\documentclass[twocolumn,prb,notitlepage,floats,superscriptaddress,amsmath,amssymb]{revtex4-2}

\usepackage{float}
\usepackage[normalem]{ulem}
\usepackage[euler]{textgreek}
\usepackage{mathrsfs}
\usepackage[T1]{fontenc}
\usepackage{bm}
\usepackage{graphicx}
\usepackage{natbib}
\usepackage{amsmath}
\usepackage{textcomp}
\usepackage[dvipsnames]{xcolor}
\definecolor{MMgreen}{RGB}{0,128,0}
\usepackage{hyperref} 
\hypersetup{colorlinks=true,citecolor=blue, filecolor=blue ,linkcolor=blue , urlcolor=blue, pdftex}
\usepackage{sidecap}
\usepackage[caption=false]{subfig}

\usepackage{multirow}

\def \FUW{Institute of Experimental Physics, Faculty of Physics, University of Warsaw, 02-093 Warsaw, Poland}
\def \Watanabe{Research Center for Electronic and Optical Materials, National Institute for Materials Science, 1-1 Namiki, Tsukuba 305-0044, Japan}
\def \Taniguchi{Research Center for Materials Nanoarchitectonics, National Institute for Materials Science,  1-1 Namiki, Tsukuba 305-0044, Japan}
\def \LNCMI{Laboratoire National des Champs Magnétiques Intenses, CNRS-UGA-UPS-INSA-EMFL, 38042 Grenoble, France}
\def \Brno{Central European Institute of Technology, Brno University of Technology, Brno, Czech 61200, Republic}
\def \Centera{CENTERA Laboratories, Institute of High Pressure Physics, Polish Academy of Sciences, 01-142 Warsaw, Poland}

\begin{document}

\title{Raman scattering excitation in monolayers of semiconducting transition metal dichalcogenides}

\author{M. Zinkiewicz}
\email{malgorzata.zinkiewicz@fuw.edu.pl}
\affiliation{\FUW}
\author{M.~Grzeszczyk}
\affiliation{\FUW}
\author{T. Kazimierczuk}
\affiliation{\FUW}
\author{M. Bartos}
\affiliation{\Brno}
\author{K. Nogajewski}
\affiliation{\FUW}
\author{W. Pacuski}
\affiliation{\FUW}
\author{K.~Watanabe}
\affiliation{\Watanabe}
\author{T. Taniguchi}
\affiliation{\Taniguchi}
\author{A. Wysmo\l{}ek}
\affiliation{\FUW}
\author{P. Kossacki}
\affiliation{\FUW}
\author{M. Potemski}
\affiliation{\FUW}
\affiliation{\LNCMI}
\affiliation{\Centera}
\author{A. Babi\'nski}
\affiliation{\FUW}
\author{M. R. Molas}
\email{maciej.molas@fuw.edu.pl}
\affiliation{\FUW}

\begin{abstract}

Raman scattering excitation (RSE) is an experimental technique in which the spectrum is made up by sweeping the excitation energy when the detection energy is fixed. 
We study the low-temperature ($T$=5~K) RSE spectra measured on four high quality monolayers (ML) of semiconducting transition metal dichalcogenides (S-TMDs), $i.e.$ MoS$_2$, MoSe$_2$, WS$_2$, and WSe$_2$, encapsulated in hexagonal BN.
The outgoing resonant conditions of Raman scattering reveal an extraordinary intensity enhancement of the phonon modes, which results in extremely rich RSE spectra.
The obtained spectra are composed not only of Raman-active peaks, $i.e.$ in-plane E$'$ and out-of-plane A$'_1$, but the appearance of 1$^{st}$, 2$^{nd}$, and higher-order phonon modes is recognised.
The intensity profiles of the A$'_1$ modes in the investigated MLs resemble the emissions due to neutral excitons measured in the corresponding PL spectra for the outgoing type of resonant Raman scattering conditions.
Furthermore, for the WSe$_2$ ML, the A$'_1$ mode was observed when the incoming light was in resonance with the neutral exciton line.
The strength of the exciton-phonon coupling (EPC) in S-TMD MLs strongly depends on the type of their ground excitonic state, $i.e.$ bright or dark, resulting in different shapes of the RSE spectra.
Our results demonstrate that RSE spectroscopy is a powerful technique for studying EPC in S-TMD MLs.
\end{abstract}

\maketitle

\section{Introduction \label{sec:Intro}}
The electron-phonon coupling is, in addition to the Coulomb interaction, one of the fundamental interactions between quasiparticles in solids~\cite{Vogl1980}.
It plays an important role in a variety of physical phenomena, in particular, low-energy electronic excitations can be strongly modified by coupling to lattice vibrations, which influences $e.g.$ their transport~\cite{Huewe2015} and thermodynamic~\cite{Zhou2020} properties.

Semiconducting transition metal dichalcogenides (\mbox{S-TMDs}) based on molybdenum and tungsten, $i.e.$ MoS$_2$, MoSe$_2$, MoTe$_2$, WS$_2$, and WSe$_2$, are the most well-known representatives of van der Waals (vdW) materials~\cite{Koperski2017, Wang2018}. 
Due to the very strong absorption and direct energy band gap in the monolayer (ML) limit, in recent years this class of materials has become of great interest from both research~\cite{Mak2010, Cadiz2017, Koperski2017, Robert2017} and development point of view~\cite{Duan2015,Ye2016}.
The photoluminescence (PL) signal of S-TMDs is caused mainly by excitonic effects, even at room temperature, due to the large excitonic binding energy at the level of hundreds of meV~\cite{Stier2016, Goryca2019, Molas2019Spectrum}. 
It arises from the reduced dimensionality of the material and limited dielectric screening of the environment~\cite{Zhu2015}. 
Constant progress in sample preparation, particularly the encapsulation of MLs in thin layers of hexagonal BN (hBN)~\cite{Cadiz2017}, leads to narrowing of the observed emission lines to the limit of a few meV, which opens the possibility of studying a variety of individual excitonic complexes associated with both bright and dark states~\cite{Courtade2017, Li2018, Chen2018, Barbone2018, Paur2019, Liu2019, Li2019Trion, Li2019Replica, Li2019Momentum, Molas2019Dark, LiuGate, LiuValley, Liu2020, He2020valley, Robert2020, Lu2020, Zinkiewicz2020, Robert2021PRL, Zinkiewicz2021, Zinkiewicz2022}.

\begin{center}
    \begin{figure*}[th!]
		\centering
		\includegraphics[width=1\linewidth]{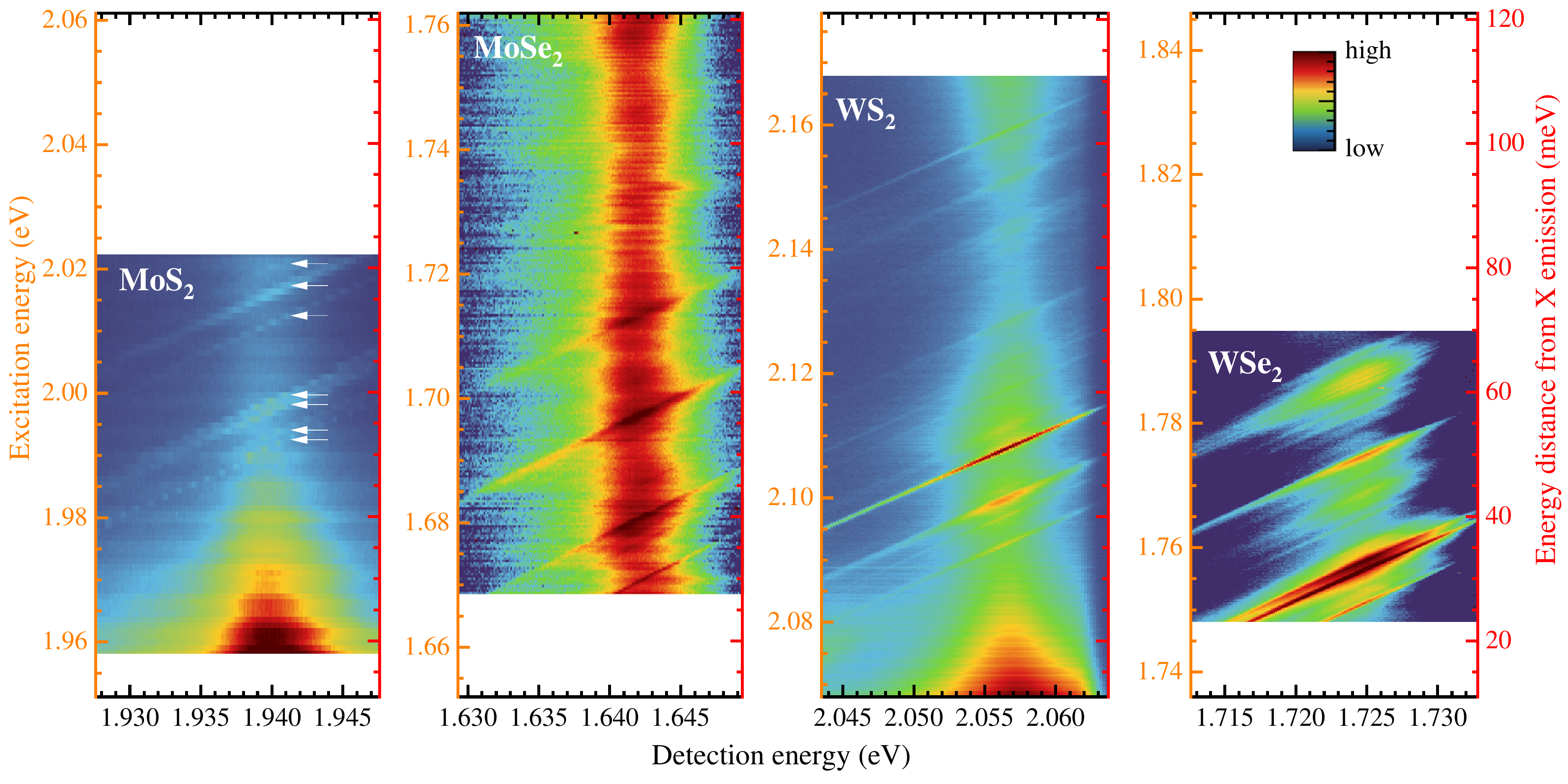}%
		\caption{False-colour maps of optical response of the MoS$_2$, MoSe$_2$, WS$_2$, and WSe$_2$ MLs encapsulated in hBN measured at low temperature ($T$=5~K) under excitation of tuneable lasers (excitation power $\sim$150~$\mu$W). 
        The maps are vertically aligned with respect to the energy distance from the energy of the neutral exciton (X) emission in these materials, which is marked on the red right-hand energy scales.
        The colour scales have been normalised to the maximum intensity.
        The detection energies of the maps are centred around the X emission. 
        The phonon modes are visible as several narrow parallel resonances passing diagonally across the maps.
        For the MoS$_2$ ML, white arrows point out lines which are investigated in the following.}
		\label{fig:RSE_map}
    \end{figure*}	
\end{center}

Raman scattering excitation (RSE) experiment performed on S-TMD MLs was proposed a few years ago in Ref.~\citenum{Molas2017_scirep} as a powerful technique to investigate the interaction between different excitonic complexes and phonons, $i.e.$ exciton-phonon coupling (EPC). 
This approach is analogous to the PL excitation (PLE) method, in which the detected spectra are measured as a function of the excitation energy, while the detection energy is fixed. 
In the RSE experiment, the detection energy is set to be equal to the emission of different excitonic complexes, while the excitation energy is larger than the former one by only a few dozen meV.
This results in the outgoing resonant conditions of Raman scattering (RS)~\cite{Feng1989}, while phonon modes can be observed in the vicinity of the excitonic emission, allowing one to study the exciton-phonon interaction in a given material.
RSE spectra were previously reported for bare MLs of MoSe$_2$~\cite{Chow2017, Shree2018} and WS$_2$~\cite{Molas2017_scirep} exfoliated on Si/SiO$_2$ substrates.  
In the case of the MoSe$_2$ MLs~\cite{Chow2017, Shree2018}, the RSE spectrum was dominated by several phonon replicas of the LA mode.
In contrast, for the WS$_2$ ML~\cite{Molas2017_scirep}, a very rich Raman spectrum was presented with numerous phonon modes originating from the edge of the Brillouin zone (BZ) as well as multiphonon modes.
Knowing that S-TMD MLs are organised into two subgroups, $i.e.$ $bright$ and $darkish$, due to the type of the ground exciton state (bright and dark, respectively)~\cite{Molas2017, Molas2019Dark, Robert2020, Zinkiewicz2020}, their low-temperature ($T$$\sim$4.2 -- 20~K) PL spectra display completely different complexity.
The high quality of the MLs embedded between the hBN flakes, accompanied by the division of the MLs into two subgroups, has motivated us to conduct a comprehensive study devoted to the EPC in such S-TMD MLs.

In this work, we use the RSE technique to investigate the exciton-phonon interaction in four high-quality samples consisting of the MoS$_2$, MoSe$_2$, WS$_2$, and WSe$_2$ MLs encapsulated in hBN flakes. 
We observe the intensity enhancements of Raman modes, while their emission energies match the emission energy of the corresponding neutral exciton (X).
The measured RSE spectra are composed of many peaks that can be attributed not only from the BZ centre ($\Gamma$ points) but also from other points in the BZ ($e.g.$ M points), which are followed by lines arising from multiphonon processes.
In the case of outgoing resonance conditions with the X emission, the intensity profiles of the A$'_1$ modes in the studied MLs resemble the emission lines of the corresponding X emission.
The A$'_1$ modes' enhancements are extraordinarily strong for the WS$_2$ and WSe$_2$ MLs, but the corresponding increases of the A$'_1$ peaks in the MoS$_2$ and MoSe$_2$ MLs are significantly smaller.
Instead, for the Mo-based MLs, the modes involving LA phonons from the edge of the BZ, such as 2LA, 3LA, and so on, are greatly enhanced.
Moreover, we observe that the A$'_1$ intensity is also significantly enhanced when the excitation energy is in the vicinity of the X emission, which leads to the incoming resonance condition.
We propose that the difference in the obtained RSE measured on S-TMD MLs at low temperature can be understood in terms of the division of MLs into $bright$ and $darkish$ subgroups.

\section{Raman scattering excitation spectra \label{results}}

Fig.~\ref{fig:RSE_map} presents false colour maps of the low-temperature ($T$=5~K) emission intensities collected for the MoS$_2$, MoSe$_2$, WS$_2$, and WSe$_2$ MLs encapsulated in hBN flakes. 
The vertical position of a given map has been shifted to reflect the relative energy of the neutral exciton (X) emission.
The assignment of the X line as arising from the neutral A exciton is straightforward and consistent with many other studies on these MLs~\cite{Cadiz2017, Molas2019Spectrum, Molas2019Dark, Zinkiewicz2020, Zinkiewicz2021}.
We focus on the X lines, but we are aware that the low-temperature PL spectra of the investigated MLs are composed of several additional emission lines (see, for example, Fig.~\ref{fig:inout}), particularly due to charged excitons, dark complexes, and the corresponding phonon replicas~\cite{Grzeszczyk2021, Klein2022, Park2022, Smolenski2019, Barbone2018, Paur2019, He2020valley, LiuValley, Zinkiewicz2021, Rodek2023}.

It can be seen that the line shape of the detected signal signiﬁcantly depends on the excitation energy.
Several parallel narrow lines superimposed on neutral exciton emissions are clearly observed in Fig.~\ref{fig:RSE_map}.
These sharp lines follow the tuned excitation energy, which points out the Raman scattering as their origin and are examined in detail in the following.
Moreover, for the MoS$_2$ and WS$_2$ MLs, the X emissions are strongly enhanced with decreasing excitation energies towards the respective X energies.
The observed enhancement of the X intensity can be described in terms of extremely efficient formation of neutral excitons at larger $k$-vectors due to the near-resonant excitation.
Note that the laser line used in experiments is narrow enough to investigate individual narrow peaks, but the excitation range of measurements is limited by the spectral range of the lasers. 
Moreover, the dye laser applied for MoS$_2$ measurements has a limited tuning accuracy.
This leads to the observation of the narrow lines only at specific excitation and detection energies, in contrast to almost continuous linear evolutions present for other MLs; see Fig.~\ref{fig:RSE_map}.
As is seen in the figure, the intensities of the Raman modes strongly depend on the material.
For Mo-based MLs, $i.e.$ MoS$_2$ and MoS$_2$, the intensity enhancement of different phonon modes is similar and relatively small compared to the X intensity.
In contrast, extraordinary increases in the selected Raman peaks accompanied by a large number of smaller modes are observed in the spectra measured on the WS$_2$ and WSe$_2$ MLs.

\begin{center}
	\begin{figure}[t!]
		\centering
		\includegraphics[width=1\linewidth]{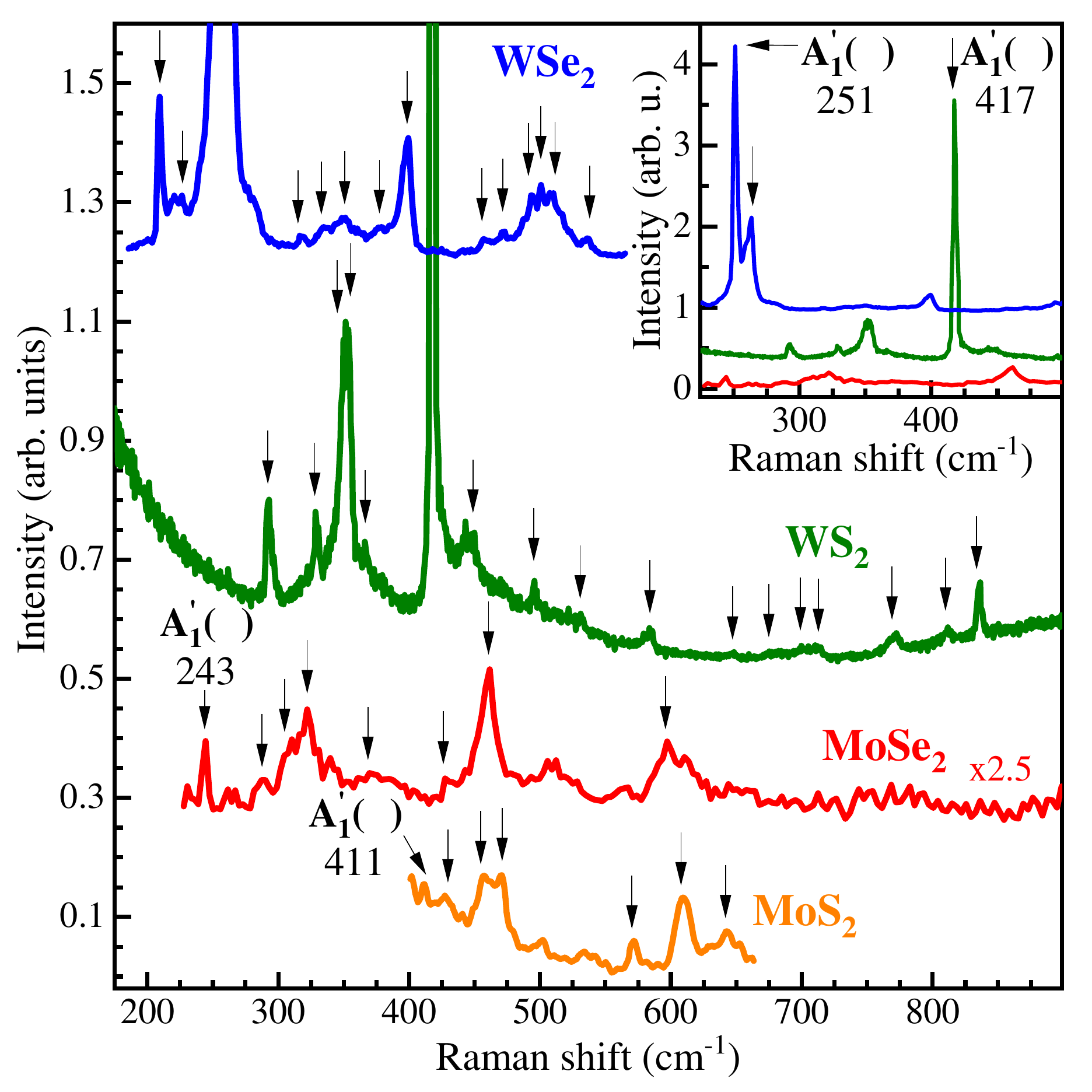}%
		\caption{Raman scattering excitation (RSE) spectra detected at the energies of the X line measured on the MoS$_2$, MoSe$_2$, WS$_2$, and WSe$_2$ MLs encapsulated in hBN.
        The vertical scale has been adjusted to make the low-intensity peaks visible, while the inset shows the RSE spectra with the most pronounced peaks.
        Note that the RSE spectrum of the MoSe$_2$ ML was multiplied by a factor of 2.5 for clarity.
        The black arrows denote the peaks which are identified, and their assignment is presented in Table S1 in the ESI.
        The A$'_1$ peaks are marked with their energy in cm$^{-1}$.}
		\label{fig:RSE_spectra}
	\end{figure}
\end{center}

In order to investigate in detail the apparent phonon modes in the aforementioned optical responses of the MLs, the RSE spectra are plotted in Fig.~\ref{fig:RSE_spectra}.
The RSE spectra were obtained by fixing the detection energy at the emission energy of the X line, while the excitation energy was being tuned.
Note that the horizontal energy scale in the figure corresponds to the relative distance between the excitation and X energies, and is given in cm$^{-1}$, which is a typical unit for RS experiments and represents the so-called Raman shift.
As can be seen in the inset of Fig.~\ref{fig:RSE_spectra}, the intensity of the A$'_1$ modes in the W-based MLs is extraordinarily high and surpasses the X emission several times.
In contrast, the phonon modes related to the longitudinal acoustic branch, $i.e.$ 2LA, 3LA, \dots, are enhanced for the Mo-based MLs, while the corresponding intensities of the A$'_1$ modes are much smaller. 
All of the presented RSE spectra are very rich, consisting of many phonon modes.
This confirms the resonant excitation conditions of RS, since the corresponding non-resonant Raman spectra are composed of only two Raman-active modes in the backscattering geometry of the experiment, $i.e.$ A$'_1$ and E$'$~\cite{Zhang_review}, whereas RS spectra become especially rich under resonant excitation conditions~\cite{Golasa2014}.
Using the previous results presented in the literature~\cite{Golasa2014, Shi2016, Soubelet2016, Carvalho2017, Molas2017_scirep, Zinkiewicz2019}, we have identified almost all observed peaks, marked in Fig.~\ref{fig:RSE_spectra} with vertical black arrows.
Their energies and assignments to the respective phonons are summarised in the Electronic Supporting Information (ESI).
Besides the two Raman-active modes from the $\Gamma$ point of the BZ, the majority of the observed peaks were ascribed to phonons from the M points, which are located at the edge of the hexagonal BZ of the MLs in the middle between the K$^+$ and K$^-$ points. 
For example, we have identified peaks related to combinations of all three branches of acoustic phonons ($e.g.$ LA, ZA, TA) as well as their higher orders ($e.g.$ 2LA, 3LA,\dots); see the ESI for details.
Their observation can be associated with the resonant excitation of RS, which results in the appearance of Raman inactive momentum-conserving combinations of acoustic modes from the edge of the BZ~\cite{Golasa2014, Molas2019kwas, Bhatnagar2022}.
The absence of features due to a single LA phonon in the RS spectra is in line with the momentum conservation rule. 
This confirms the high quality of the investigated structures, as their appearance in the RS spectra heralds disorder in the system~\cite{GolasaGold}.

We have also investigated the RSE spectrum measured on the MoSe$_2$ ML grown on hBN flake using the molecular beam epitaxy (MBE) technique~\cite{Pacuski2020}. 
The ML was also covered with a thin hBN flake using mechanical exfoliation and dry transfer technique.
The spectra measured for epitaxial layers exhibit peaks at energies similar to those in the case of exfoliated layers, but the relative intensity of various peaks differs. 
Particularly pronounced for epitaxial layers are LA replicas, which are observed from 2$^{nd}$ to 12$^{th}$ order, see ESI for details.

\section{Enhancement profiles of A$'_1$ modes due to outgoing resonance\label{results}}

\begin{center}
    \begin{figure*}[t]
		\centering
		\includegraphics[width=1\linewidth]{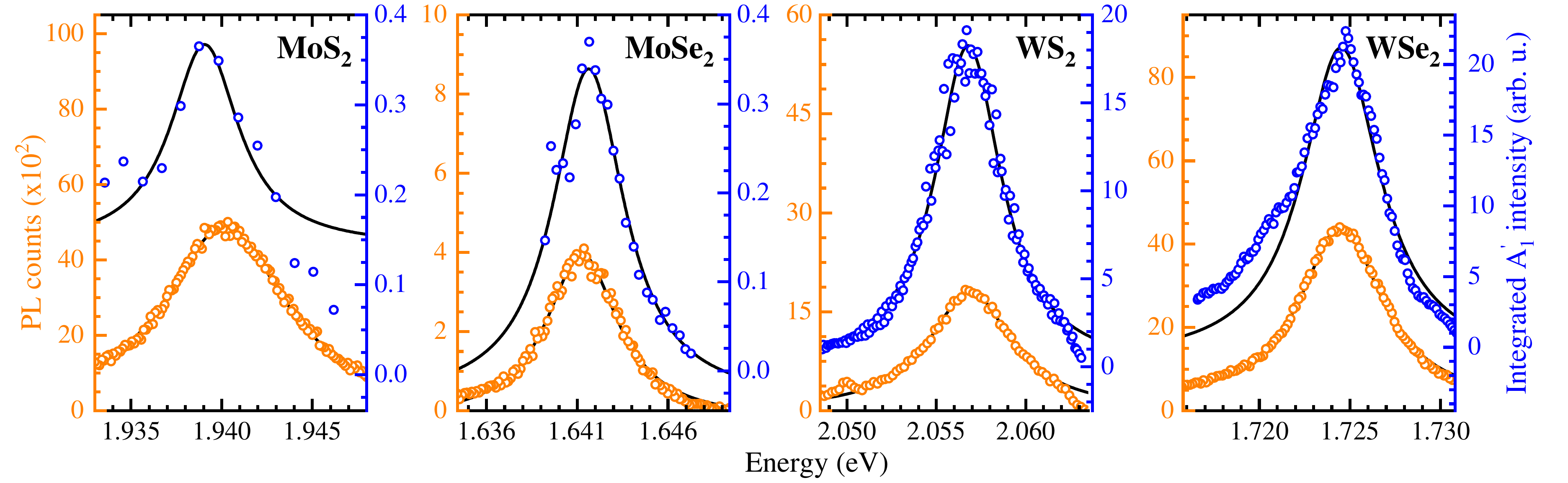}%
		\caption{Comparison of (orange points) PL emissions of the X lines and (blue points) integrated intensities of the A$'_1$ peaks measured on the MoS$_2$, MoSe$_2$, WS$_2$, and WSe$_2$ MLs encapsulated in hBN.
        Solid black curves represent the corresponding fitting using Lorentzian function.}
		\label{fig:profile}
	\end{figure*}
\end{center}

The resonant enhancement of the phonon modes presented above originates from the so-called outgoing resonance, which requires that the resonant excitation energy must be equal to the sum of the phonon and exciton energies~\cite{Zinkiewicz2019}.
To verify this hypothesis we compare the PL emissions of the X lines and the integrated intensities of the A$'_1$ peaks in Fig.~\ref{fig:profile}. 
The choice of A$'_1$ modes is motivated by their substantial intensities compared to the remaining phonon modes, see Fig.~\ref{fig:RSE_spectra}.
Note that we also analyse some other selected modes, $e.g.$ 2ZA, 2LA, $etc.$, in the ESI, which show a similar behaviour as the one found for A$'_1$.
As can be seen in Fig.~\ref{fig:profile}, the data were fitted using Lorentzian functions, which nicely reproduce the profiles shown for both the X line and the A$'_1$ peak.
Furthermore, the parameters of the Lorentzian profiles fitted for the X lines and the A$'_1$ peaks are summarised in Table~\ref{tab:profile}.
Except for the results obtained for the MoS$_2$ ML, the extracted profiles for the A$'_1$ peaks, described by their energies and linewidths, follow the shape of the X emission line.
For the MoS$_2$ ML, the observed small discrepancy in the energies of the X and A$'_1$ peaks of about 1~meV may be explained in terms of the small energy resolution of the laser used in this case.
Nevertheless, the results obtained confirm that the enhancements of the A$'_1$ peaks in the S-TMD MLs are due to the outgoing resonance.

\begin{table}[t!]
\centering
\caption{Summary of the obtained fitting parameters of the X emission lines and A$'_1$ profiles using Lorentzian functions, shown in Fig.~\ref{fig:profile}. $x_c$ and $w$ represent fitted energy position and the full width at half maximum (FWHM, linewidth), respectively.}
    \label{tab:profile}
\begin{tabular}{cc|c|c|c|c|}
   &       & \textbf{MoS$_2$} & \textbf{MoSe$_2$} & \textbf{WS$_2$} & \textbf{WSe$_2$} \\\cline{1-6} 

\multirow{4}{*}{\textbf{X}}  & \multirow{2}{*}{$x_c$ (eV)}  &   1.9402   &        1.6413       &    2.0569         &   1.7273     \\      
                                 & &  $\pm$0.0002        &      $\pm$0.0002     &    $\pm$0.0002    &   $\pm$0.0001              \\ \cline{2-6} 
                                 & \multirow{2}{*}{$w$ (meV)} &      8.3    &           4.9        &       5.7          &    5.5  \\   
                                 &  &  $\pm$0.1    &    $\pm$0.1     &      $\pm$0.1          &   $\pm$0.1       \\\cline{1-6}

\multirow{4}{*}{\textbf{A}$\mathbf{'_1}$} & \multirow{2}{*}{$x_c$ (eV)} & 1.9391 & 1.6416  & 2.0567  &  1.7245 \\ 
                                 & &  $\pm$0.0009 & $\pm$0.0005  & $\pm$0.0002  &  $\pm$0.0005          \\ \cline{2-6}  
                                 & \multirow{2}{*}{$w$ (meV)} &  4.6  &  4.8  &   4.5  &   5.3     \\
                                 &  &  $\pm$1.6  &  $\pm$0.5  &   $\pm$0.1    &   $\pm$0.4     \\

\end{tabular}

\end{table}

\section{Outgoing versus incoming resonance \label{results}}

\begin{center}
    \begin{figure}[b!]
		\centering
		\includegraphics[width=1\linewidth]{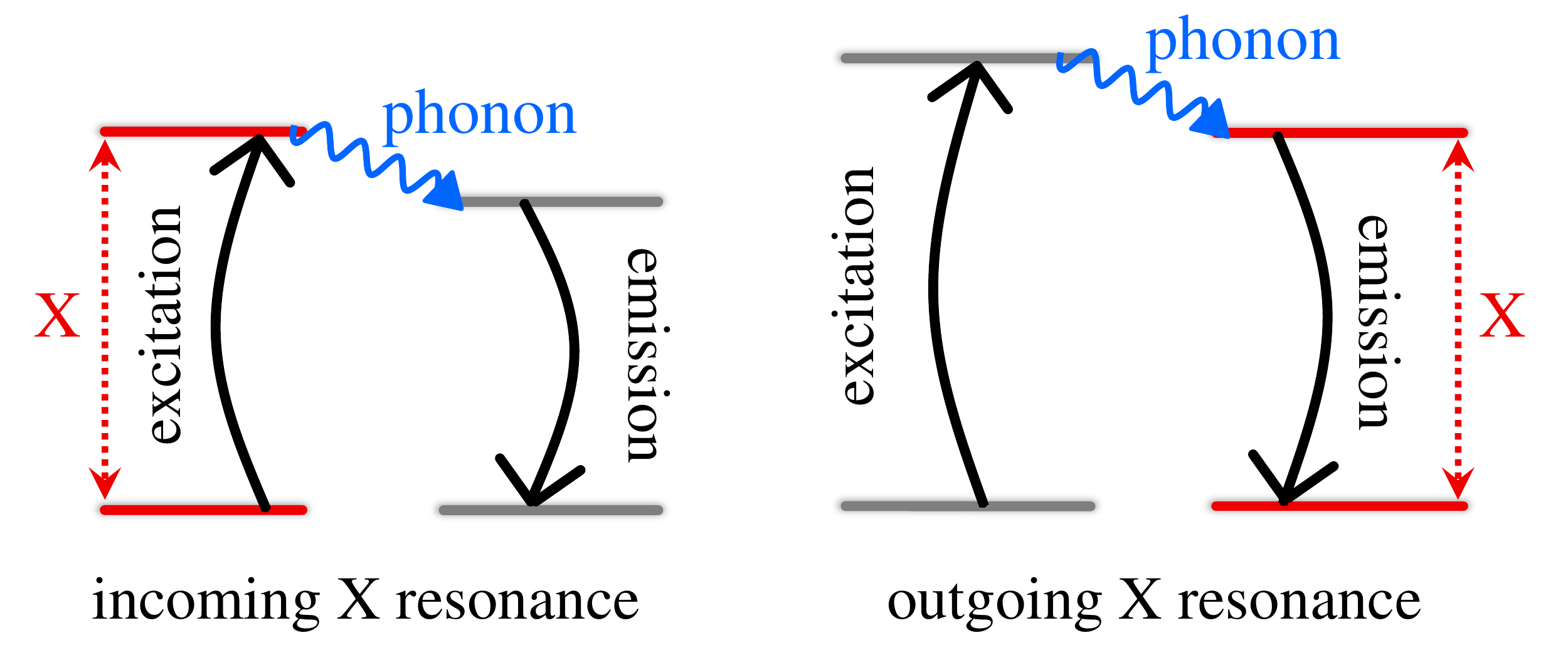}%
		\caption{Schematic illustration of the (left panel) incoming and (right panel) outgoing excitation resonance with the excitonic transition, denoted as X.}
		\label{fig:schemat}
	\end{figure}
\end{center}

Fig.~\ref{fig:schemat} shows a schematic illustration of the incoming and outgoing excitation resonance with the excitonic transition (X). 
The resonance type depends on the resonant conditions between the incident or scattered light with the X energy, while the energy difference between the excitation and the emission amounts to the phonon energy.
The results presented so far are associated with the outgoing X resonance, which occurs when the scattered-photon energy is equal to that of the optical X transition.
In contrast, the incoming X resonance takes place when the incident photon energy equals the energy of an optical X transition.

\begin{center}
    \begin{figure}[t]
		\centering
		\includegraphics[width=1\linewidth]{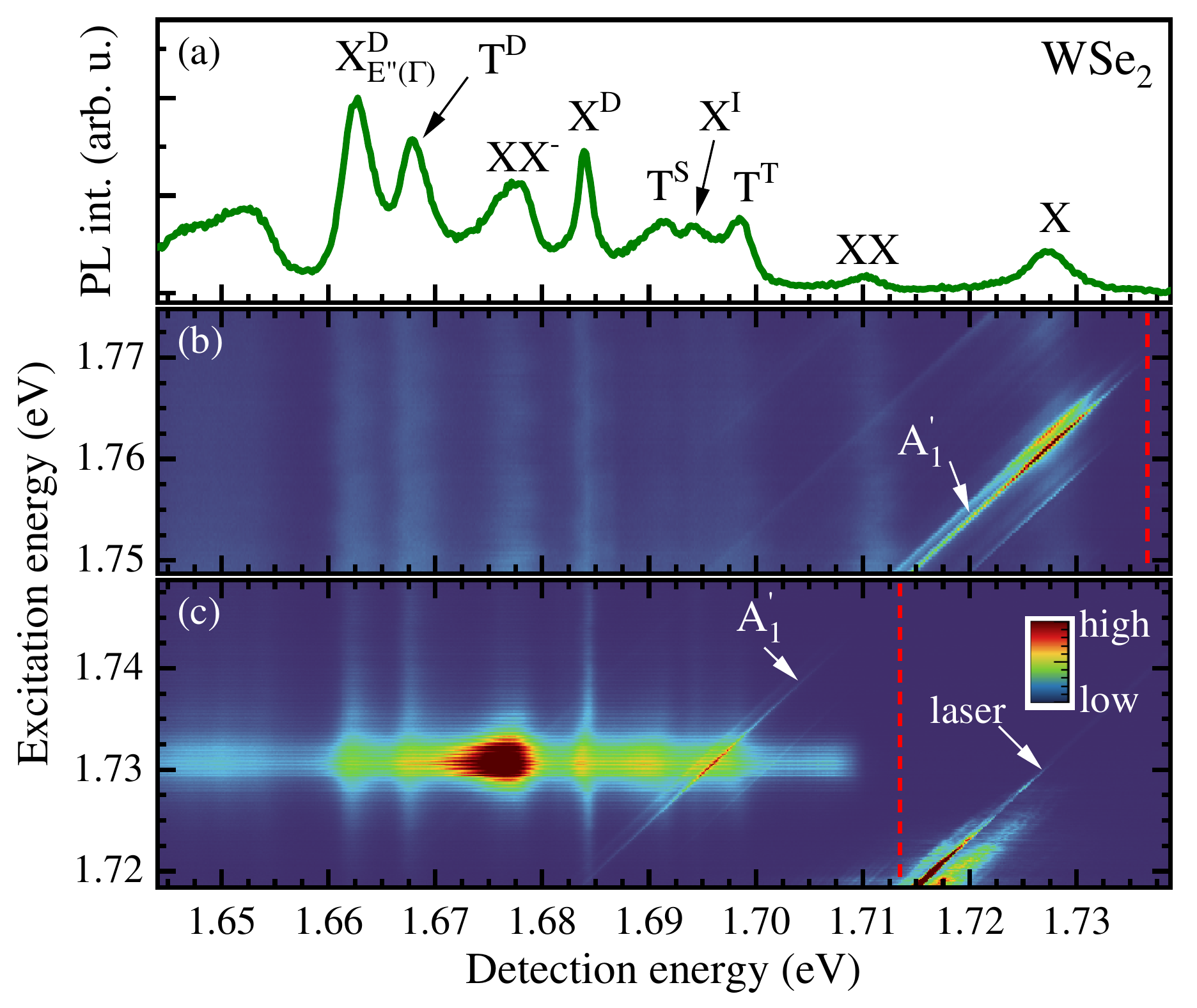}%
		\caption{(a) Low-temperature ($T$=5 K) PL spectrum of the WSe$_2$ ML encapsulated in hBN flakes measured using excitation energy of 2.4 eV and power of 150 $\mu$W. 
        The lines' assignments are as follows:  X -- neutral exciton; XX -- neutral biexciton; T$^\textrm{S}$ and T$^\textrm{T}$ -- singlet (intravalley) and triplet (intervalley) negatively charged excitons, respectively; X$^\textrm{I}$ and X$^\textrm{D}$ -- momentum- and spin-forbidden neutral dark excitons, respectively; XX$^-$ -- negatively charged biexciton; T$^\textrm{D}$ -- negatively charged dark exciton (dark trion); X$^\textrm{D}_{\textrm{E}"(\Gamma)}$ -- phonon replica of the X$^\textrm{D}$ complex.
        (b) and (c) False-colour maps of low-temperature ($T$=5~K) optical response of the WSe$_2$ MLs encapsulated in hBN under excitation of tuneable-energy lasers (excitation power $\sim$10~$\mu$W) measured in the vicinity of the outgoing and incoming resonance conditions of Raman scattering with the X line, respectively.
        The colour scales in panels (b) and (c) have been normalised to the maximum intensity.
        The vertical red dashed lines denote the edge of the used longpass filters.}
		\label{fig:inout}
	\end{figure}
\end{center}

To study the possibility of achieving the incoming resonant conditions in S-TMD MLs, we measured the optical response of the WSe$_2$ ML in both the outgoing and incoming resonance conditions, see Fig.~\ref{fig:inout}.
The low-temperature PL spectrum of the WSe$_2$ ML encapsulated in hBN flakes, presented in panel (a) of the figure, displays several emission lines with a characteristic pattern similar to that previously reported in several works on WSe$_2$ MLs embedded in between hBN flakes~\cite{Courtade2017, Li2018, Chen2018, Barbone2018, Paur2019, Liu2019, Li2019Trion, Li2019Replica, Li2019Momentum, Molas2019Dark, LiuGate, LiuValley, Liu2020, He2020valley, Robert2020, Lu2020, Zinkiewicz2020, Robert2021PRL, Zinkiewicz2021, Zinkiewicz2022}.
Panel (b) and (c) display correspondingly the optical response of the WSe$_2$ ML in the outgoing and incoming resonances with the neutral exciton emission.
As for the outgoing conditions, the observed results are analogous to those shown in Fig.~\ref{fig:RSE_map}, the incoming resonance leads to the two prominent effects.
The first one, seen as a significant enhancement of all the emission lines, is associated with the resonant excitation to the X transition.
Due to these conditions, the shape of the PL spectrum is strongly modified by an enormous increase in the emission intensity of the negatively charged biexciton (XX$^-$), as the nonradiative recombination processes are strongly suppressed.
The second effect is associated with the observation of several narrow peaks, parallel to the excitation laser, which are superimposed on the enhanced emission lines below the X peak.
These peaks can be ascribed to phonon modes, whose intensities are enlarged due to the incoming resonance.
As in the outgoing case, the increase in the intensity depends on the phonon symmetry, resulting in the most pronounced A$'_1$ peak.

\begin{center}
    \begin{figure}[t]
		\centering
		\includegraphics[width=1\linewidth]{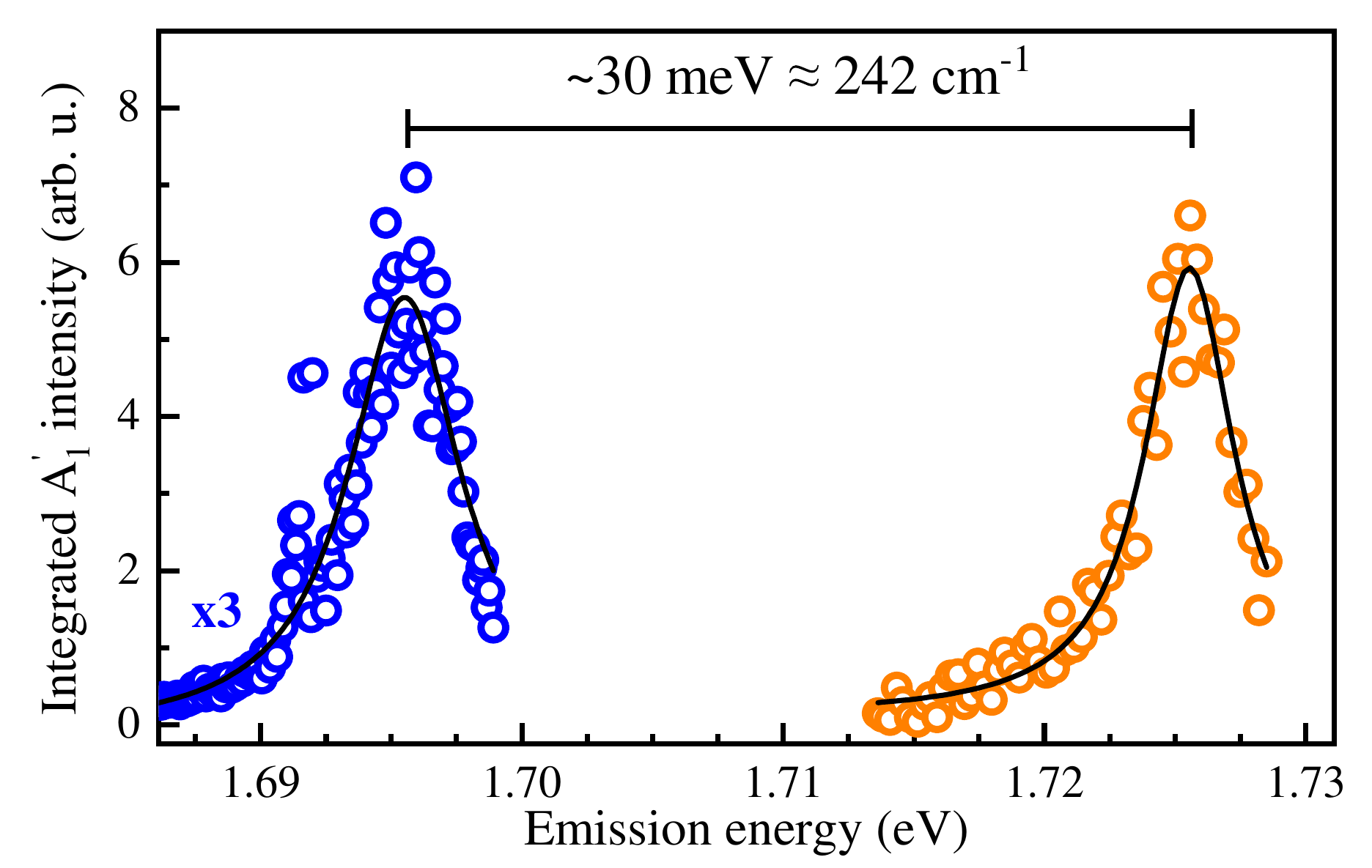}%
		\caption{Intensity profiles of the A$'_1$ mode measured on the WSe$_2$ ML encapsulated in hBN flakes.
        Blue and orange circles correspond to the incoming and outgoing resonance conditions of the Raman scattering with the X line.}
		\label{fig:out_PL}
	\end{figure}
\end{center}

Alike in Fig.~\ref{fig:profile}, we analyse the intensity profiles of the A$'_1$ mode measured on the WSe$_2$ ML under the incoming and outgoing resonance conditions of the Raman scattering with the X line, see Fig.~\ref{fig:out_PL}.
It can be seen that the A$'_1$ evolutions can be nicely described by Lorentzian functions with similar linewidths, while their intensities differ considerably.
The enhancement of the A$'_1$ intensity is about 3 times larger in the outgoing resonance as compared to the incoming one, which points out the different strength of the exciton-phonon coupling in these two regimes.
For a given phonon mode, the energy separation between the outgoing and incoming resonances with a particular transition should be equal to this phonon energy.
The obtained energy separation for the A$'_1$ mode presented in Fig.~\ref{fig:out_PL} is of about 30~meV.
This value is equivalent of 242$\pm$1~cm$^{-1}$, which is very close to the A$'_1$ wavenumber from Fig.~\ref{fig:RSE_spectra} (~$\sim$251~cm$^{-1}$).

\section{Discussion}	
In the literature on the subject, a physical picture has been so far drawn that the exciton-phonon coupling in S-TMD MLs can be understood considering the symmetries of phonon modes with respect to the symmetries of orbitals associated with the involved transitions (excitons)~\cite{Carvalho2015}.
For example, it has been established that the A$'_1$ mode is enhanced when the excitation laser is in resonance with A and B excitons in S-TMDs, while the E$'$ intensity is increased under resonanant conditions matching the energy of C excitons~\cite{Carvalho2015, Soubelet2016}.
Our results, shown in Fig.~\ref{fig:RSE_spectra}, demonstrate a substantial difference in the intensity and complexity of the RSE spectra measured on the Mo- and W-based MLs.
While the A$'_1$ modes are extraordinarily enhanced for the WS$_2$ and WSe$_2$ MLs leading to the extremely rich RSE spectra (up to 16 identified phonon modes in the WS$_2$ ML), the RSE spectra of the MoS$_2$ and MoSe$_2$ MLs are dominated by the modes involving acoustic phonons, $e.g.$ 2LA, 3LA, \dots.
In our opinion, this difference can be understood in terms of the excitonic states in the vicinity of their band gaps. 
Although S-TMD MLs share a very similar band structure, $i.e.$ they are direct band-gap semiconductors with the minima (maxima) of the conduction (valence) band located at the K$^+$ and K$^-$ points of the BZ~\cite{Koperski2017,Wang2018}, the arrangement of the optically active transitions in the vicinity of their band gaps is different.
A strong spin–orbit coupling results in spin-split and spin-polarised subbands in both the valence band (VB) and the conduction band (CB).
Consequently, MLs of S-TMD are organised into two subgroups, $i.e.$ $bright$ and $darkish$, due to the type of the ground exciton state (bright and dark, respectively)~\cite{Molas2017}. 
In bright MLs, the excitonic recombination between the top VB and the bottom CB is optically active (bright), while the opposite happens darkish MLs.
Nowadays, it is well established that for MLs encapsulated in hBN flakes, MoSe$_2$ is bright, while MoS$_2$, WS$_2$, and WSe$_2$ MLs are darkish (see \cite{Robert2017, Molas2019Dark, Zinkiewicz2020, Robert2020}).
It means that the observed difference in the intensity of the phonon modes in the RSE spectra, shown in Fig.~\ref{fig:RSE_spectra}, may coincide with a particular alignment of bright and dark states in S-TMD MLs.

The substantial enhancement of the LA phonons from the M points in the MoSe$_2$ has been explained by the efficient phonon-assisted scattering occurring between the bright X exciton and the dark indirect exciton (IX) formed by an electron and a hole located at the Q and K points, respectively (see Ref.~\cite{Chow2017} for details).
Simultaneously, the small increase of the A$'_1$($\Gamma$) intensity in this material suggests that the strength of the EPC is not remarkable.
In contrast, the extraordinarily high intensity of the Raman-active A$'_1$($\Gamma$) phonon modes (does not require any additional scattering between different points in the BZ) in the W-based MLs resulting in the very rich Raman spectra can not be described using the same approach, see Fig.~\ref{fig:RSE_spectra}.
Particularly, similar results obtained for the WS$_2$ and WSe$_2$ MLs may suggest the involvement of the ground dark excitons in the efficient EPC in these materials.
In the following, we speculate on the possible mechanisms, which may affect the EPC in these $"darkish"$ materials. 
As the ground exciton state in W-based MLs is dark, the intensity of the bright exciton emission might be controlled by concurrent processes of the radiative recombination (emission of photons) and the relaxation process to the dark exciton. 
It is known that the relative intensity of the bright and dark excitons in WSe$_2$ is a function of temperature~\cite{Zhang2015} with a significant quenching of the X emission at low temperature.  
Consequently, a large reservoir of dark excitons in W-based MLs is formed at low temperature~\cite{Robert2017, Molas2019Dark, Zinkiewicz2020}, while the analogous ones will be absent in bright MLs (ground exciton state is bright).
It suggests that the strength of the EPC is associated with not only the symmetries of both phonons and electronic bands~\cite{Carvalho2015}, but may be affected by other phenomena.
Our results indicate that the order of excitonic states, $i.e.$ bright versus dark, should also be included in the analysis of the EPC in S-TMD MLs.
The most complex are the results obtained for the MoS$_2$ ML, which are very similar to the ones of the MoSe$_2$ ML, $i.e.$ the LA phonons from the M points are significantly enhanced.
It is known that the MoS$_2$ MLs encapsulated in hBN flakes exhibit dual character.
The theoretically predicted structure of the CB and VB yields optical activity of the energetically lowest transition, while including the excitonic effects leads to the dark ground excitonic state~\cite{Robert2020, Grzeszczyk2021}.
This may indicate that the EPC in the MoS$_2$ ML is more associated with its electronic band structure than with the excitonic one.
Summarising, the aforementioned phenomena, which may be responsible for the EPC in S-TMD MLs, requires solid theoretical analysis, which is beyond the scope of our experimental work.

\section{Methods \label{methods}}
The studied samples were composed of four S-TMD MLs, $i.e.$ MoS$_2$, MoSe$_2$, WS$_2$ and WSe$_2$, encapsulated in hBN flakes and supported by a bare Si substrate. 
The structures were obtained by two-stage polydimethylsiloxane (PDMS)-based~\cite{Gomez2014} mechanical exfoliation of bulk crystals of S-TMDs and hBN. 
A bottom layer of hBN in heterostructures was created in the course of non-deterministic exfoliation to achieve the highest quality. 
The assembly of heterostructures was realised via successive dry transfers of a ML and capping hBN flake from PDMS stamps onto the bottom hBN layer.

Optical measurements were performed at low temperature ($T$=5~K) using typical setups for the PL and PLE experiments. 
The investigated sample was placed on a cold finger in a continuous-flow cryostat mounted on $x$-$y$ manual positioners. 
The non-resonant PL measurements were carried out using 514.5~nm (2.41~eV) and 532~nm (2.33~eV) radiations from the continuous wave Ar$^+$ and Nd:YAG lasers, respectively. 
To study the optical response of a ML as a function of excitation energy, $i.e.$ to measure PLE spectra, two types of tuneable lasers were used: dye lasers based on Rhodamine 6G and DCM, and a Ti:Sapphire laser. 
The excitation light was focused by means of a 50× long-working distance objective that produced a spot of about 1 $\mu$m diameter. 
The signal was collected via the same microscope objective, sent through a monochromator, and then detected by a charge-coupled device (CCD) camera.
Measurements of the low-energy part of the RSE spectra, $i.e.$ from around 15~meV from the laser line, were carried out using ultra steep edge long-pass filters mounted in front of the spectrometer.

\section{Summary}
We have presented the investigation of the exciton-phonon interaction in four high-quality samples consisting of MoS$_2$, MoSe$_2$, WS$_2$, and WSe$_2$ MLs encapsulated in hBN flakes. 
By sweeping the excitation energy for a fixed value of the detection energy, we have observed an astonishing amplification of phonon-modes' intensity, while the detection energy was in resonance with neutral exciton emission. 
Due to this phenomenon, the measured RSE spectra are composed of many phonon modes originating not only from the BZ centre ($\Gamma$ points) but also from others ($e.g.$ M points), which are followed by lines due to multiphonon processes.
For the outgoing X resonance, we have found that the intensity profiles of the A$'_1$ modes in the studied MLs resemble the emission lines of the corresponding X emission.
The A$'_1$ enhancements are extraordinarily strong for the WS$_2$ and WSe$_2$ MLs, but an analogous effect in the MoS$_2$ and MoSe$_2$ MLs are significantly smaller.
For the Mo-based MLs, the modes involving acoustic phonons, such as 2LA and 3LA, are significantly enhanced.
Moreover, we have observed that the A$'_1$ intensity is also greatly strengthened since the excitation energy is in the vicinity of the X emission, leading to the incoming resonance condition.
We have proposed that the difference in the obtained RSE measured on S-TMD MLs at low temperature can be understood in terms of the division of MLs into $bright$ and $darkish$ subgroups.
These results shine new light on the exciton-phonon interaction in the MLs of S-TMDs, pointing out that not only the symmetries of phonons and excitons play an important role in this process, but also the type of their ground excitonic states.

\section*{Acknowledgements}
The work was supported by the National Science Centre, Poland (grants no. 2017/27/B/ST3/00205, 2018/31/B/ST3/02111, and 2021/41/B/ST3/04183), EU Graphene Flagship Project, and the CNRS via IRP "2DM" project. 
M.B. acknowledges support from the the Grant Agency of the Czech Republic under grant no. 23-05578S.
K.W. and T.T. acknowledge support from the JSPS KAKENHI (Grant Numbers 21H05233 and 23H02052) and World Premier International Research Center Initiative (WPI), MEXT, Japan.

\bibliographystyle{apsrev4-2}
\bibliography{RSE_ML.bib}

\newpage
\onecolumngrid
\setcounter{figure}{0}
\setcounter{section}{0}
\renewcommand{\thefigure}{S\arabic{figure}}
\renewcommand{\thetable}{S\arabic{table}}
\renewcommand{\thesection}{S\arabic{section}}
	\begin{center}
	{\large{{\bf  \textsc{Electronic Supplementary Information:}} \\ Raman scattering excitation in monolayers of semiconducting transition metal dichalcogenides}}
\end{center}

\title{Electronic Supplementary Information:\\ }

\begin{center}
This electronic supplementary information provides: \ref{Raman_RSE} attribution of the phonon modes observed in the RSE spectra of S-TMD MLs, \ref{MBE} Raman scattering spectra measured on the MoSe$_2$ ML grown using molecular beam epitaxy, and \ref{Rprofile} intensity profiles of phonon modes in the outgoing resonance with the neutral exciton. 
\end{center}


\section{A\lowercase{ttribution of the phonon modes observed in the} RSE \lowercase{spectra of} S-TMD ML\lowercase{s} \label{Raman_RSE}}

Fig.~2 in the main text shows the Raman scattering excitation (RSE) spectra measured on four monolayers (MLs) of semiconducting transition metal dichalcogenides (S-TMDs), $i.e.$ MoS$_2$, MoSe$_2$, WS$_2$, and WSe$_2$, encapsulated in hexagonal BN (hBN) and detected at the energies of the neutral exciton (X) line.
The peaks denoted with vertical black arrows point to the identified ones.
Table~\ref{tab:RSE} summarises the identified Raman peaks with their energies in cm$^{-1}$ and assignments. 
The identification of the observed peaks has been done based on previous measurements presented in Refs.~\cite{Golasa2014, Carvalho2017} for MoS$_2$, Ref.~\cite{Soubelet2016} for MoSe$_2$, Refs.~\cite{Molas2017_scirep, Zinkiewicz2019} for WS$_2$, and Ref.~\cite{Shi2016} for WSe$_2$. 
Most of the observed phonon modes come from the $M$ point of the hexagonal Brillouin zone (BZ) of the MLs, which is located at the edge of the BZ in the middle between the K$^+$ and K$^-$ points. 
We have also identified phonon modes from the centre of the BZ, $i.e.$ $\Gamma$ point, as well as from the K points, located at the corners of the BZ.

\begin{table*}[h!]
\caption {Energies in cm$^{-1}$ and attributions of the phonon modes observed in the RSE spectra detected at the energies of the X line measured on the MoS$_2$, MoSe$_2$, WS$_2$, and WSe$_2$ MLs encapsulated in hBN presented in Fig.~2 in the main text. Designated phonons come from the M point in the Brillouin zone, otherwise their origin is indicated in brackets.}
\label{tab:RSE}
\begin{tabular}{lcccccccccc}
\multicolumn{2}{c}{MoS$_2$}       &      &  
\multicolumn{2}{c}{MoSe$_2$}      &      & 
\multicolumn{2}{c}{WS$_2$}        &      & 
\multicolumn{2}{c}{WSe$_2$}     
\\
\cline{1-2} \cline{4-5} \cline{7-8} \cline{10-11}
\multicolumn{1}{c|}{$\omega$ {(}cm$^{-1}${)}} & Mode     &    & 
\multicolumn{1}{c|}{$\omega$ {(}cm$^{-1}${)}} & Mode     &    & 
\multicolumn{1}{c|}{$\omega$ {(}cm$^{-1}${)}} & Mode     &    & 
\multicolumn{1}{c|}{$\omega$ {(}cm$^{-1}${)}} & Mode       
\rule[-2mm]{0mm}{6mm}
\\ 
\cline{1-2} \cline{4-5} \cline{7-8} \cline{10-11}
\multicolumn{1}{c|}{411} & {A'$_{1}$($\Gamma$)} & &
\multicolumn{1}{c|}{243} & {A'$_{1}$($\Gamma$)} & & 
\multicolumn{1}{c|}{293} & 2ZA                  & & 
\multicolumn{1}{c|}{209} & {TA+ZA}    
\rule[-1mm]{0mm}{5mm}
\\
\multicolumn{1}{c|}{428} & LA(K)+TA(K) & &
\multicolumn{1}{c|}{289} & E'($\Gamma$)  & &
\multicolumn{1}{c|}{382} & LA+ZA         & &
\multicolumn{1}{c|}{224} & 2ZA/E"
\rule[-1mm]{0mm}{5mm}
\\
\multicolumn{1}{c|}{456} & 2LA(K)   & &
\multicolumn{1}{c|}{310} & 2LA      & &
\multicolumn{1}{c|}{350} & 2LA      & & 
\multicolumn{1}{c|}{251} & {A'$_{1}$($\Gamma$)}
\rule[-1mm]{0mm}{5mm}
\\
\multicolumn{1}{c|}{471} & 2LA & &
\multicolumn{1}{c|}{322} & A'$_{1}$+TA   &  & 
\multicolumn{1}{c|}{353} & E'($\Gamma$)  &  & 
\multicolumn{1}{c|}{264} & 2LA      
\rule[-1mm]{0mm}{5mm}
\\
\multicolumn{1}{c|}{571} & {2E''($\Gamma$)} & &
\multicolumn{1}{c|}{372} & A'$_{1}$+LA     & & 
\multicolumn{1}{c|}{366} & E'              & & 
\multicolumn{1}{c|}{318} & E''+TA   
\rule[-1mm]{0mm}{5mm}
\\
\multicolumn{1}{c|}{609} & E''+LA        & &
\multicolumn{1}{c|}{427} & E'+LA        & & 
\multicolumn{1}{c|}{417} & A'$_{1}$($\Gamma$) & & 
\multicolumn{1}{c|}{335} & 3ZA
\rule[-1mm]{0mm}{5mm}
\\
\multicolumn{1}{c|}{643} & {A'$_{1}$+LA} & &
\multicolumn{1}{c|}{462} & 3LA           & & 
\multicolumn{1}{c|}{443} &               & & 
\multicolumn{1}{c|}{350} & E'+TA       
\rule[-1mm]{0mm}{5mm}
\\
& & &
\multicolumn{1}{c|}{597} & 4LA     &   & 
\multicolumn{1}{c|}{496} & 2LA+ZA  &   &
\multicolumn{1}{c|}{378} & E'+LA    
\rule[-1mm]{0mm}{5mm}
\\
& & &
\multicolumn{1}{c|}{768} & 5LA    &   & 
\multicolumn{1}{c|}{531} & 3LA    &   & 
\multicolumn{1}{c|}{399} & 3LA  
\rule[-1mm]{0mm}{5mm}
\\
& & & & 
\multicolumn{1}{l}{} & 
\multicolumn{1}{l}{} & 
\multicolumn{1}{c|}{584} & A'$_{1}$+LA   &  &
\multicolumn{1}{c|}{457} & E'+2TA      
\rule[-1mm]{0mm}{5mm}
\\
& & & & 
\multicolumn{1}{l}{}  &
\multicolumn{1}{l}{}  &
\multicolumn{1}{c|}{648} & 2LA+2ZA    &  & 
\multicolumn{1}{c|}{472} & E'+ZA   
\rule[-1mm]{0mm}{5mm}
\\
& & &  & 
\multicolumn{1}{l}{}   & 
\multicolumn{1}{l}{}   & 
\multicolumn{1}{c|}{680} & 3LA+ZA     &   &
\multicolumn{1}{c|}{493} & E'+2LA    
\rule[-1mm]{0mm}{5mm}
\\
& & &  & 
\multicolumn{1}{l}{}   & 
\multicolumn{1}{l}{}   & 
\multicolumn{1}{c|}{700}  & 4LA        &  & 
\multicolumn{1}{c|}{501}  & 2A'$_{1}$($\Gamma$)    
\rule[-1mm]{0mm}{5mm}
\\
& & &  & 
\multicolumn{1}{l}{}    &
\multicolumn{1}{l}{}    &
\multicolumn{1}{c|}{710}  & 2E'($\Gamma$)  &   &
\multicolumn{1}{c|}{509}  & E'+2LA     
\rule[-1mm]{0mm}{5mm}
\\
& & &  &   &   & 
\multicolumn{1}{c|}{772}  & E'($\Gamma$)+A'$_{1}$($\Gamma$)  &  & 
\multicolumn{1}{c|}{539}  & 2A'$_{1}$   
\rule[-1mm]{0mm}{5mm}
\\
& & &  &   &   &
\multicolumn{1}{c|}{812}  & 2A'$_{1}$  &  &  &   
\rule[-1mm]{0mm}{5mm}
\\
& & &  &   &   &
\multicolumn{1}{c|}{836}  & 2A'$_{1}$($\Gamma$)   &   &   &            
\rule[-1mm]{0mm}{5mm}
\end{tabular}
\end{table*}

\newpage
\section{RSE \lowercase{spectra measured on} MBE\lowercase{-grown} M\lowercase{o}S\lowercase{e}$_2$ ML \label{MBE}}

\begin{center}
    \begin{figure*}[h!]
		\centering
		\includegraphics[width=1\linewidth]{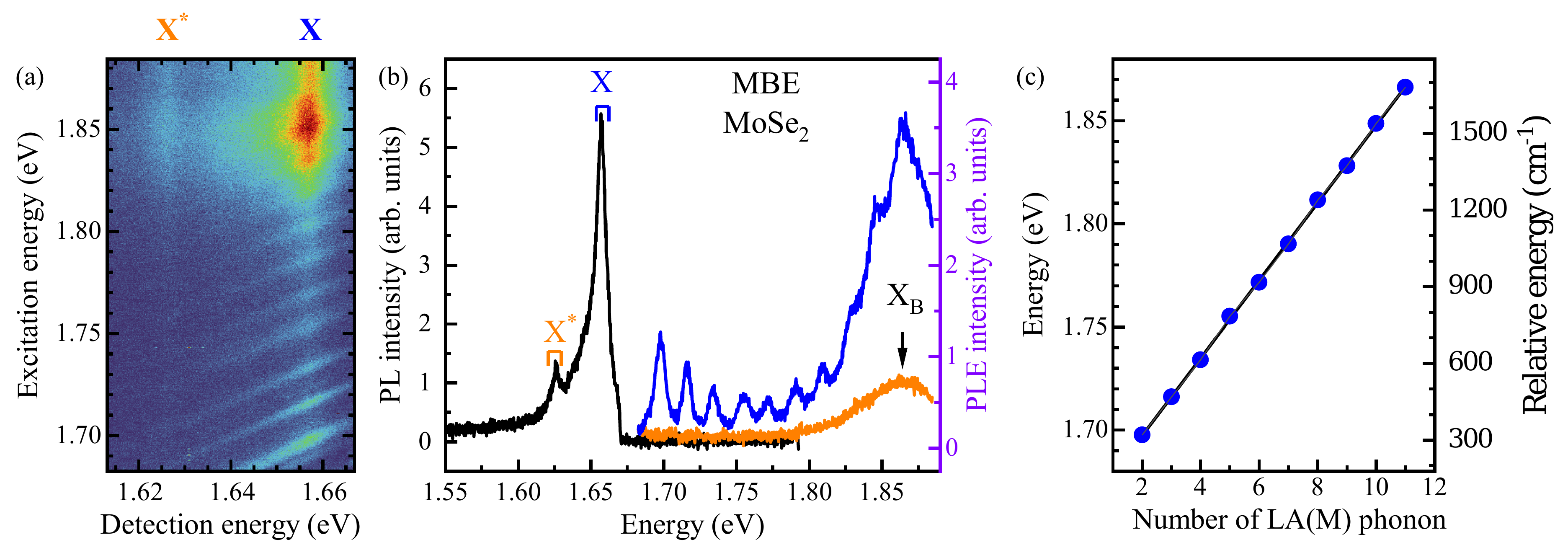}%
		\caption{(a) False-colour maps of the optical response of a MoSe$_2$ ML grown on an hBN flake using the MBE technique measured at low temperature ($T$=5~K) under the excitation of supercontinuum tuneable lasers (excitation power $\sim$50~$\mu$W). 
        (b) PL and PL excitation (PLE) spectra measured on the MoSe$_2$ ML grown on the hBN flake using the MBE technique.
        (c) The energies of the LA(M) phonon modes as a function of their number.}
		\label{fig:fig_2_SI}
    \end{figure*}	
\end{center}

As the observed RSE signal measured on the MoSe$_2$ ML was much weaker than for other samples (see Fig.~2 in the main text), we decided to perform an analogous study on a monolayer of different origin.
We had the opportunity to investigate the RSE spectrum on a MoSe$_2$ ML grown on an hBN flake using the molecular beam epitaxy (MBE) technique~\cite{Pacuski2020}. 
To perform the RSE experiment in the broad spectral range, we chose a supercontinuum laser combined with a monochromator as a tuneable excitation source.
It provides monochromatic excitation light with the linewidth of around 2~nm, characterised by a broad energy tunability.
Fig.~\ref{fig:fig_2_SI}(a) presents a false colour map of the low-temperature ($T$=5~K) emission intensity collected for the MBE-grown MoSe$_2$ ML covered with an hBN flake. 
The PL spectra of the ML are composed of two narrow emission lines, which can be clearly attributed to the neutral (X) and charged (X$^*$) excitons~\cite{Robert2020}.
As in the main text, several parallel narrow lines superimposed on the X emission are clearly observed, while the analogous signal is absent on the X$^*$ emission line.
The RSE spectra detected on the X and X$^*$ lines are shown in Fig.~\ref{fig:fig_2_SI}(b)
The spectrum related to the X transition presents multiple narrow lines, whose linewidths are determined by the linewidth of the excitation laser.
There is also a broader resonance attributed to the B excitons at the K points of the ML's BZ~\cite{AroraMoSe2}.
For the X$^*$ case, only the enhancement due to the B transition is observed. 
This points to the difference in the exciton-phonon coupling for excitonic complexes characterised by a different net charge, reported earlier in Refs.~\cite{Molas2017_scirep, Chow2017, Shree2018}.
We have ascribed the narrow peaks seen in the RSE spectrum of the X line to the modes due to the consecutive emissions of the LA phonons from the M point of the BZ.
We plot the energies of the LA modes as a function of their number in Fig.~\ref{fig:fig_2_SI}(c).
We assign the LA phonon replicas from the 2$^{nd}$ to 12$^{th}$ order. 
A simple linear fit to the data confirms the origin of these peaks, as the energy difference between them has been determined to $\sim$19~meV$\approx$152~cm$^{-1}$, which is very close to the energy of the LA(M) phonon obtained from Table~\ref{tab:RSE} ($\sim$155~cm$^{-1}$). 
In particular, the enhancement of the LA phonons seen in Fig.~\ref{fig:fig_2_SI} is much larger compared to the results presented in Fig.~1 and 2 in the main text.
This difference may be assigned to the different structures of the investigated MoSe$_2$ MLs.
While the ML investigated in the main text was a single flake that was exfoliated from a bulk crystal, the ML grown using the MBE technique has almost a continuous form with only small isolated parts of the bottom hBN flake that had not been covered by MoSe$_2$ and only small inclusions of the MoSe$_2$ bilayer, as reported in Ref.~\cite{Pacuski2020}.
We believe that both of these characteristics may favour phonon scattering, which is reflected in a stronger RSE signal in the MBE-grown MoSe$_2$ ML as compared to the exfoliated one.

\newpage
\section{I\lowercase{ntensity profiles of phonon modes in the outgoing} X \lowercase{resonance} \label{Rprofile}}

\begin{center}
    \begin{figure*}[h!]
		\centering
		\includegraphics[width=1\linewidth]{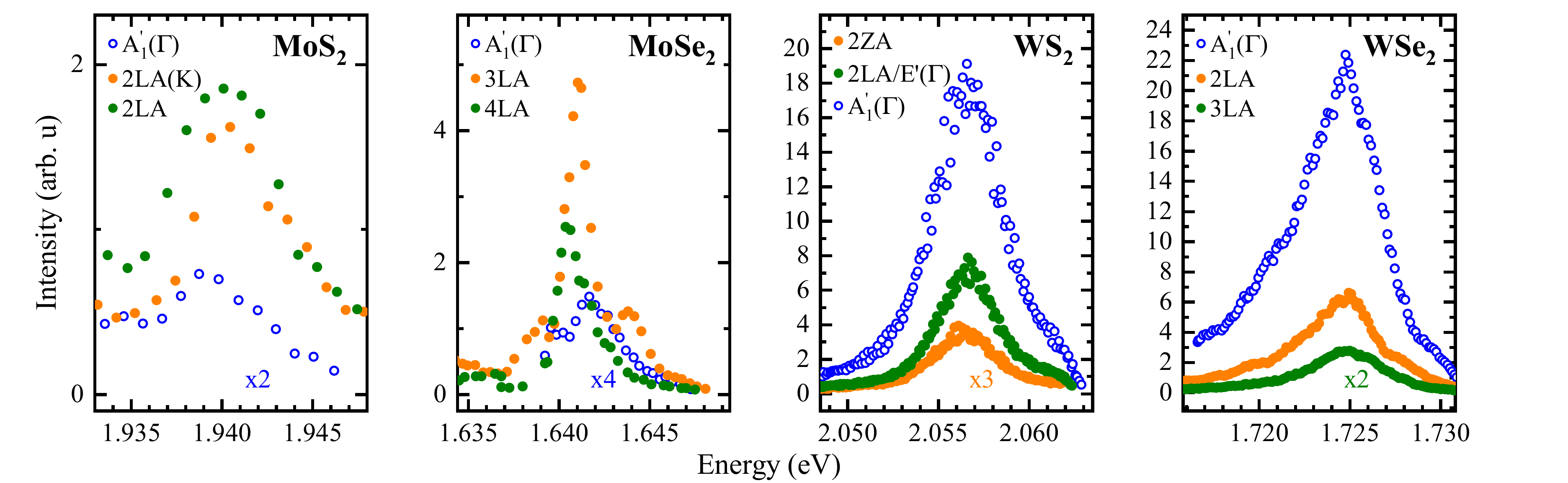}%
		\caption{Integrated intensities of the phonon peaks measured on the MoS$_2$, MoSe$_2$, WS$_2$, and WSe$_2$ MLs encapsulated in hBN plotted as a function of their emission energy. 
        The blue points correspond to the A$'_1$ modes, already introduced in the Fig.~3 in the main text. 
        Some of the data are multiplied by the indicated factors for clarity. }
		\label{fig:fig_1_SI}
    \end{figure*}	
\end{center}

To perform a more comprehensive analysis of the outgoing X resonance, we compared the enhancement profiles of different phonons.
Fig.~\ref{fig:fig_1_SI} presents the extracted intensity profiles for the most intense phonon modes seen in the RSE spectra measured on the MoS$_2$, MoSe$_2$, WS$_2$, and WSe$_2$ MLs encapsulated in hBN (see Fig. 2 in the main text).
Note that the blue circles represent the A$'_1$ profiles, which were already introduced in Fig.~3 in the main text.  
As can be seen, all of the intensity profiles presented have maxima located in the vicinity of the A$'_1$ profiles, which indicates the interaction of these phonons with the neutral exciton.
For the W-based ML, the A$'_1$ mode is the most intensive and is a few times larger than other analysed modes related to acoustic phonons.
In the case of Mo-based MLs, the situation is opposite. 
The greatest enhancement is present for the acoustic modes, $e.g.$ in-plane longitudinal (2LA, 3LA, \ldots) and out-of-plane (ZA).
For example, the intensity of the 3LA mode seen for the MoSe$_2$ ML is more than one order of magnitude higher than that for the A$'_1$ one. 
The origin of these different enhancements of phonon modes in Mo- and W-based MLs is considered in Discussion section of the main text.

\end{document}